\documentclass[aps,pra,twocolumn,10pt,longbibliography,nobibnotes]{revtex4-2}
\usepackage{amsmath,amssymb,graphicx,mathtools,dsfont,courier}
\usepackage[T1]{fontenc}
\usepackage{xcolor}
\usepackage{ulem}
\begin{document}

\newcommand{\bra}[1]    {\langle #1|}
\newcommand{\ket}[1]    {|#1 \rangle}
\newcommand{\ketbra}[2]{|#1\rangle\!\langle#2|}
\newcommand{\braket}[2]{\langle#1|#2\rangle}
\newcommand{\tr}[1]    {{\rm Tr}[ #1 ]}
\newcommand{\trr}[2]    {{\rm Tr}[ #1 ]_{\overline{#2}}}
\newcommand{\titr}[1]    {\widetilde{{\rm Tr}}\left[ #1 \right]}
\newcommand{\avs}[1]    {\langle #1 \rangle}
\newcommand{\modsq}[1]    {| #1 |^2}
\newcommand{\0}    {\ket{\vec 0}}
\newcommand{\1}    {\ket{\vec 1}}
\newcommand{\dt}    {\delta\theta}
\newcommand{\I}    {\mathcal  I_{an}^{(q)}}
\newcommand{\Id}[1]    {\mathcal  I_q\left[#1\right]}
\newcommand{\Ic}[1]    {\mathcal  I_{\hat A}\left[#1\right]}
\newcommand{\C}    {\hat{\mathcal C}(t)}
\newcommand{\Cd}    {\hat{\mathcal C}^\dagger(t)}
\newcommand{\re}[1]    {\texttt{Re}\left[#1\right]}
\newcommand{\im}[1]    {\texttt{Im}\left[#1\right]}
\newcommand{\red}[1]{\textcolor{red}{#1}}

\title{Transfer of quantum-enhanced information through a many-body system}

\author{Piotr Wysocki$^{1}$, Marcin P{\l}odzie\'n$^{2,3}$ and Jan Chwede\'nczuk$^{1}$}
\affiliation{$^{1}$ Faculty of Physics, University of Warsaw, ulica Pasteura 5, 02-093 Warszawa, Poland\\
    $^{2}$Qilimanjaro Quantum Tech, Carrer de Veneçuela 74, 08019
Barcelona, Spain\\
  $^{3}$ICFO-Institut de Ciencies Fotoniques, The Barcelona Institute of Science and Technology, 08860 Castelldefels (Barcelona), Spain 
  }

\begin{abstract}
  Forthcoming quantum devices will require high-fidelity information transfer across a many-body system. 
  We formulate the criterion for lossless signal propagation and show that a single qubit can play the role of an antenna, collecting large amounts of information from a complex system.
  We derive the condition under which the antenna, far from the source and embedded in a many-body interacting medium, can still collect the complete information. 
  A striking feature of this setup is that a single qubit antenna can receive even the full signal amplified by the entanglement of the source. 
  As a consequence, the recovery of this information can be performed with simple single-qubit operations on the antenna (which we fully characterize) rather than with multi-qubit measurements of the source. 
  Finally, we discuss the control of the system parameters 
  necessary for lossless signal propagation. A method discussed here could improve the precision of quantum devices and simplify metrological protocols. 
\end{abstract}

\maketitle

\section{Introduction}

The array of castles built in the valley of the Adige River in northern Italy used bonfires to exchange warnings of the approaching enemy. 
The structures formed a ``conveyor belt'' for information that was sent along the river.
This information-oriented view of complex systems is central to both classical~\cite{Shannon1948} and quantum~\cite{Ingarden1976} technologies~\cite{dittrich2022information}.
Quantum metrology relies on the fact that some entangled states can store large amounts of information about the quantity 
being measured~\cite{pezze2018quantum,horodecki2009entanglement,brunner2014bell,horodecki2024multipartiteentanglement, RevModPhys.91.025001,Zou2021, PhysRevLett.126.210506,Zhang2024,Scandi_2025}.
Another example is the quantum-based communication which  uses the Quantum State Transfer protocol~\cite{QST2014},
extensively studied in the context of many-body quantum systems, in particular spin-$1/2$ chains
~\cite{Eckert2007,DiFranco2008,  Markiewicz2009, DiFranco2010,Wang2012,Vinet2012,Apollaro2015,Zwick2015,Vogell2017,Agundez2017,Huang2018,Iversen2020,Huang2021,Xu2023,Jameson2023,Yue2024}. 

In this work, we show that a collection of qubits can form a quantum equivalent of this centuries-old conveyor belt
allowing the lossless transfer of information on some parameter $\theta$ between its distant parts. 
Our workhorse is quantum Fisher information (QFI), which is the maximum amount of 
information about the parameter $\theta$ that can be extracted from a density matrix $\hat\varrho(\theta)$  using any quantum measurements~\cite{braunstein1994statistical},
\begin{align}\label{eq.qfi}
  \Id{\hat\varrho}=2\sum_{i,j}\frac{\modsq{\bra{\psi_j}\dot{\hat\varrho}\ket{\psi_i}}}{p_i+p_j},
\end{align}
where $\ket{\psi_{i,j}}$ and $p_{i/j}$ are its eigenstates and eigenvalues, while the dot denotes the derivative over $\theta$. 
We show that this information can be exchanged between distant subsystems with either no loss or a small distance- and particle-independent decrement. 
We use separable and entangled states as
initial probes that collect information about $\theta$ and become a {\it source} that sends it through the system. 
The source transmits information to the antenna through the fully quantum interaction of these subsystems.
resulting in the propagation of the $\theta$-dependent signal.
We show that if the source is highly entangled, so that it 
collects an amount of information that exceeds the classical limit, all this quantum-enhanced signal can be sent without loss to just a single receiving qubit, here called an {\it antenna}. 
Hence the protocol discusses here is substantially  different from the trasfer of a full quantum state across 
a spin chain~\cite{Bose2003,Christandl2004,Osborne2004,Albanese2004,Zoller2005,Kimble2008, KAY2010,Chapman2016,Bezaz2025,khalilipourghorbaniarezoomand2024}.  
In the latter case, the state transfer utilizes the mirror symmetry of the governing Hamiltonian, focusing on the fidelity between the sent and received states.

We identify the measurements that extract the full information from the antenna and discuss 
the impact of possible experimental misalignments on
the efficiency of the protocol. Thus, by establishing the conditions under which the information 
transfer is effective, the proposed protocol could simplfy the operating principle of future quantum sensors and other non-classical devices.

\begin{figure}[t!]
    \centering
    \includegraphics[width=0.95\linewidth]{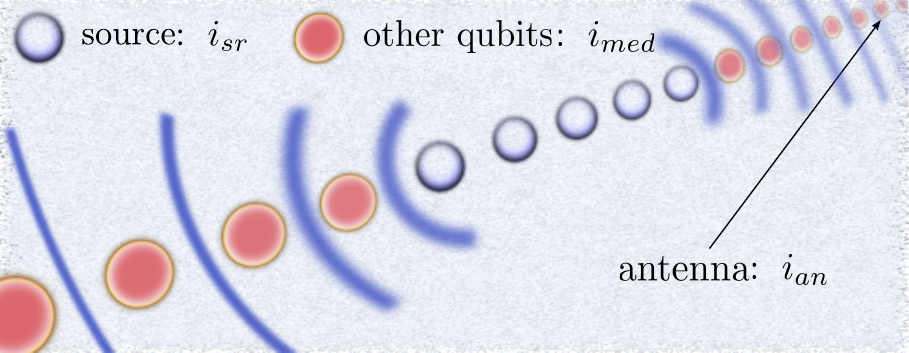}
    \caption{Schematic representation of a chain of qubits, part of which is a source of signal (blue, labelled with $i_{sr}$). 
      The remaining qubits (red) are the medium through which the information propagates ($i_{med}$)
      to reach a distant antenna ($i_{an}$).}
    \label{fig.chain}
\end{figure}

\section{Formulation of the problem}

Consider a quantum system described by a density matrix $\hat\varrho$. A part of the system, the source mentioned above, acquires information about a parameter $\theta$ via
a local Hamiltonian $\hat H_{sr}$, i.e. 
\begin{align}\label{eq.rot}
  \hat\varrho\ \longrightarrow\  \hat\varrho(\theta)=e^{-i\hat H_{sr}\theta}\hat\varrho\, e^{i\hat H_{sr}\theta}.
\end{align}
At this stage, the complete information about $\theta$, quantified by $\Id{\hat\varrho_{sr}}$, is
contained in the density operator of the source $\hat\varrho_{sr}(\theta)=\trr{\hat\varrho(\theta)}{sr}$.
The overline indicates that the trace is computed over the part of Hilbert space that is complementary to the source's degrees of freedom.

A subsequent evolution of the whole system, generated by a Hamiltonian $\hat H$, distributes the 
information about the parameter throughout the system, $\hat\varrho(\theta;t)=e^{-i\hat H t}\hat\varrho(\theta)\, e^{i\hat Ht}.$
We are interested in the amount of information that reaches another part, an antenna. In particular, we are looking for scenarios of  lossless transfer of information about the parameter $\theta$, 
$\Id{\hat\varrho_{sr}}=\Id{\hat\varrho_{an}}$,
where $\hat\varrho_{an}(\theta;t)=\trr{\hat\varrho(\theta;t)}{an}$ is the density matrix of the antenna.

\section{Illustration: spin chain}

Consider a  chain of $N$ qubits in a quantum state $\hat\varrho$, the paradigmatic platform for quantum technologies. Part of the chain, $M$ qubits,
forms the source and we label these particles with index $i_{sr}$. We call the remaining $N-M$ qubits of the array ``the medium'' and label these qubits with the index $i_{med}$.. 
A single qubit embedded in the medium is the antenna, labelled with $i_{an}$. This configuration is illustrated in Fig.~\ref{fig.chain}

The Eq.~\eqref{eq.rot} yields the $\theta$-dependent density matrix
\begin{align}\label{eq.rotated}
  \hat\varrho(\theta)=\sum_{\vec s,\vec s'=\pm1}\varrho_{\vec s,\vec s'}(\theta)\ketbra{\vec s}{\vec s'},
\end{align}
where $\ket{\vec s}=\bigotimes_{i=1}^N\ket{\pm1}^{(i)}_z$ is a product of $N$ single-qubit eigenstates of the  Pauli operators, $\hat\sigma_z^{(i)}\ket{\pm1}^{(i)}_z=\pm\ket{\pm1}^{(i)}_z$,
and the summation runs over all $2^N$ elements of the basis.

At this stage, all of the information about $\theta$ is contained in the source. To transfer this information, we will consider
the Ising Hamiltonian with zero transverse magnetic field, long-range interactions and open boundary conditions, i.e., 
\begin{align}\label{eq.ising}
  \hat H=\sum_{i>j=1}^NJ_{ij}\hat\sigma_z^{(i)}\hat\sigma_z^{(j)},
\end{align}
where $J_{ij}$ determines the strength of the coupling of qubits $i$ and $j$. The density matrix of the antenna will have the form~\cite{supp}
\begin{align}\label{eq.dens}
  \hat\varrho_{an}(\theta,t)=\left(\begin{array}{cc}
    p & a \\
    a^* & 1-p
  \end{array}\right),
\end{align}
where the probability $p$ is constant (it does not depend on either $\theta$ or $t$), while
\begin{align}\label{eq.off}
  a=\sideset{}{'}\sum_{\vec s=\pm1}\tilde\varrho_{\vec s,\vec s}(\theta)e^{-2it\sideset{}{'}\sum\limits_{i=1}^NJ_{i,i_{an}}s_i}.
\end{align}
Here the prime denotes the summation over all qubits except the antenna, which is distinguished from the surrounding medium by an index $i_{an}$~\footnote{Note that the interaction
of the source and the medium qubits does not contribute as the corresponding phase factors cancel out when the trace is calculated.}. 
Consequently, the tilde over the element of the density matrix informs that the indeces of
the antenna are fixed to $\pm1$. 
The diagonalization of this matrix gives the QFI from Eq.~\eqref{eq.qfi} equal to~\cite{supp}
\begin{align}\label{eq.full}
  \Id{\hat\varrho_{an}}=4\left(\frac{\re{\dot ae^{-i\varphi}}^2}{1-|a|^2}+\im{\dot ae^{-i\varphi}}^2\right),
\end{align}
where $\varphi=\arg(a)$.

It is reasonable to assume that the source initially forms a separable (i.e., at most classically correlated) state with the rest of the chain. Therefore the density matrix
from Eq.~\eqref{eq.rotated} takes the form
\begin{align}\label{eq.mixture}
  \hat\varrho(\theta)=\sum_ip_i\,\hat\varrho_i^{(sr)}(\theta)\otimes\hat\varrho_i^{(med)}
\end{align}
and the off-diagonal term of the antenna density matrix becomes~\cite{supp}
\begin{align}\label{eq.off.sep}
  a=\sum_{i}p_i\mathcal F_{i}^{(sr)}\,\mathcal G_{i}^{(med)}
\end{align}
where the two functions represent the coupling of the antenna to the source and to the medium in which it is embedded, 
respectively, and both take the form of Eq.~\eqref{eq.off} fed with the corresponding density matrix elements,
of either $\hat\varrho_i^{(sr)}(\theta)$ or $\hat\varrho_i^{(med)}$.

It is now clear, that---in general---the amount of information that reaches the antenna is small. This is because different phase terms of Eq.~\eqref{eq.off}
will oscillate at different rates and ``kill'' the signal. 
In principle, the statistical mixture [represented by the probability distribution $p_i$ in Eq.~\eqref{eq.mixture}] 
also degrades the information transfer. Nevertheless, there are physically sound cases where the 
signal reaches the antenna either with no loss or only slightly weaker than that sent by the source. We will now discuss two such important scenarios in detail.

\subsection{Separable state}

We start with the chain in a separable state of $N$ qubits
\begin{align}
  \ket{\psi}=\ket{+1}_x^{\otimes N}.
\end{align}
The transformation~\eqref{eq.rot}, for example taken as a rotation around the $y$ axis, acts on the  $M$ source qubits
\begin{align}\label{eq.rot.2}
  \hat H_{sr}=\frac12\sum_{i_{sr}=1}^M\hat\sigma_y^{(i_{sr})}.
\end{align}
Hence the amount of information on $\theta$ is
\begin{align}\label{eq.amount}
  \Id{\ket{\psi_{sr}}}=M.
\end{align}
A subsequent evolution~\eqref{eq.ising} gives the off-diagonal term of the antenna's density matrix in the form of Eq.~\eqref{eq.off.sep} with only single non-zero element of the sum and
\begin{subequations}
  \begin{align}
    &\mathcal F_{an}(sr)=\prod_{i_{sr}=1}^M\left[\cos(\phi_{i_{sr}})+i\sin(\theta)\sin(\phi_{i_{sr}})\right]\label{eq.f}\\
    &\mathcal G_{an}(med)=\prod_{i_{med}=1}^{\mu}\cos(\phi_{i_{med}})\label{eq.g}
  \end{align}
\end{subequations}
and $\phi_{i}=2tJ_{i,i_{an}}$.
Here $\mu=N-M-1$ is the number of qubits of the medium to which the antenna is coupled. 
Unless the phases $\phi_{i_{sr}}$ are all equal to some $\phi_1$---i.e.,  $J_{i_{sr},i_{an}}=J_1$ for all source qubits---a product of multiple functions oscillating with different frequencies 
will yield a very
small value of $\mathcal F$. Analogously, it is necessary that $\phi_{i_{med}}=\phi_2$ for all $i_{ch}$ ($J_{i_{med},i_{an}}=J_2$) to ensure that the information
transmitted to the antenna is large. Such symmetry represents
the all-to-all (ATA) coupling between the qubits
which can be realized in modern quantum simulator platforms based on Rydberg tweezer arrays~\cite{Weimer2010,Bluvstein2021,Ebadi2022,Ramette2022,Bluvstein2023,Evered2023}, 
trapped ions~\cite{Landsman2019,Joshi2020,Monroe2021,Manovitz2022,Nguyen2023,Guo2025}, or superconducting  qubits~\cite{Li2018,Song2019,Xu2020,Wu2024,PitaVidal2025}.
In addition, ATA models can effectively be simulated with short-range Hamiltonians~\cite{Gietka2021, HernndezYanes2022, Usui2024}.

Taking $\theta=0$ as the working point, the off-diagonal term becomes $a=\cos^{2M}(\phi_1)\cos^{\mu}(\phi_2)$, giving $\varphi=0$, while $\dot a$ is purely imaginary, hence Eq,~\eqref{eq.full} gives 
\begin{align}\label{eq.qfi.gen}
  \Id{\hat\varrho_{an}}=4\modsq{\dot a}=M^2\sin^2(\phi_1)\cos^{2(M-1)}(\phi_1)\cos^{2\mu}(\phi_2).
\end{align}
To maximize the information transfer, $\phi_2=m\pi$ must be satisfied with $m\in\mathbb N$. This fixes, e.g., the time as $t_m=m\pi/(2J_2)$. 
The remaining function can be maximized with respect to the free parameter $J_1$ expressed in units of $J_2$. 
If $m\pi\tilde J=\arctan((M-1)^{-1/2})+2k\pi$ with $k\in\mathbb N$ and $\tilde J=J_1/J_2$, we obtain
(for $M\gg1$)
\begin{align}\label{eq.sn}
  \Id{\hat\varrho_{an}}=\frac1eM.
\end{align}
Thus, the information decreases with respect to Eq.~\eqref{eq.amount} only by a constant prefactor, giving an almost
lossless transmission of the signal through a many-body medium.

If the source is a single qubit ($M=1$), the QFI from Eq.~\eqref{eq.qfi.gen} reads
\begin{align}\label{eq.qfi.an}
  \Id{\hat\varrho_{an}}=\sin^2(\phi_1)\cos^{2\mu}(\phi_2).
\end{align}
This can give up to $\Id{\hat\varrho_{an}}=1$, thus, if the source is only a single qubit, the information transfer can be lossless. The optimal settings are: $t_m=m\pi/J_2$ and $m\pi\tilde J=\pi/2+k\pi$, 
for example $J_1=\frac12 J_2$ for $m=1$ and $k=0$. Note that even for $M=1$, this is still a many-body system with a total of $N$ qubits.

We will now show that the transfer of maximum information coincides with the establishment of source--antenna entanglement. For this purpose, we compute the reduced two-qubit density matrix
$\hat\varrho_{sr;an}(t)$. 
The negativity of this operator~\cite{PhysRevLett.77.1413,HORODECKI19961,PhysRevLett.84.2726,PhysRevLett.84.2722} can be expressed as~\cite{supp}
\begin{align}
  \mathcal N(t)\equiv\Big|\sum_{\lambda_i<0}\lambda_i\Big|=\frac18\left|\alpha-\sqrt{\alpha^2+16\Id{\hat\varrho_{an}}}\right|.
\end{align}
The two qubits are entangled iff $\mathcal N(t)>0$. 
Here $\lambda_i$ are the (negative) eigenvalues of the partially transposed operator $\hat\varrho_{sr;an}(t)$ and $\alpha=1-\cos^{N-2}(4t)$, while $\Id{\hat\varrho_{an}}$ is given by Eq.~\eqref{eq.qfi.an}. 
For illustration, we have chosen the optimal
transfer parameters $J_1=1/2$ and $J_2=1$. At the instants when the QFI reaches the maximum $\Id{\hat\varrho_{an}}=1$, we have $\alpha=0$, which gives the maximum possible value of 
negativity, $\mathcal N(t)=1/2$, which is achievable only by the fully entangled Bell state~\cite{horodecki2009entanglement}.
Thus, the times when the complete information on $\theta$ reaches the antenna coincide with the formation of
a pure two-qubit Bell state.  
This is only possible if the other parts of the chain (i.e., the medium) are completely decoupled from this pair. Hence, the transfer of the signal to the
antenna is accompanied by its growing entanglement with the source and the uncoupling from the medium.

\subsection{Entangled state}

The most intriguing and surprising result comes from considering the source to be  initially in a Greenberger-Horne-Zeilinger (GHZ) state, which after the 
Hamiltonian-generated transformation~\eqref{eq.rot.2} reads
\begin{align}\label{eq.ghz}
  \ket{\psi_{sr}(\theta)}=\frac1{\sqrt2}\left(\ket{+1}_y^{\otimes M}+ie^{iM\theta}\ket{-1}_y^{\otimes M}\right).
\end{align}
At this stage, the information on $\theta$ is equal to
\begin{align}\label{eq.heis}
  \Id{\ket{\psi_{sr}(\theta)}}=M^2,
\end{align}
which is the Heisenberg limit~\cite{pezze2009entanglement}, the maximum amount of information that can be encoded in an $M$-qubit state by a linear (single-qubit) operation.

As before, each of the remaining chain qubits is prepared as $\ket{+1}_x$, so the full state is
\begin{align}\label{eq.init}
    \ket{\psi(\theta)}=\ket{\psi_{sr}(\theta)}\otimes\ket{+1}_x^{\otimes(N-M)}.
\end{align}
The reduced density matrix of the antenna has the form of Eq.~\eqref{eq.dens}, with $p=1/2$ and~\cite{supp}
\begin{align}\label{eq.f.2}
  \mathcal F_{an}(sr)=\cos^M(\phi_1)+i^M\sin(M\theta)\sin^M(\phi_1)
\end{align}
(assuming equal coupling of the anetnna to all source qubits).
The $\mathcal G$ remains unchanged and is equal to that of Eq.~\eqref{eq.g}. The substantial difference between Eqs~\eqref{eq.f} and~\eqref{eq.f.2} is that the phase is now $M$-times
amplified with respect to the previous case. With optimal settings as those leading to Eq.~\eqref{eq.sn} it yields
\begin{align}\label{eq.heis.2}
  \Id{\hat\varrho_{an}}=4\modsq{\dot a}=M^2,
\end{align}
it thus results in a lossless transfer of the complete information collected in an $M$-qubit GHZ state to a single-qubit antenna, see with Eq.~\eqref{eq.heis}.

\begin{figure}[t!]
    \centering
    \includegraphics[width=1.0\linewidth]{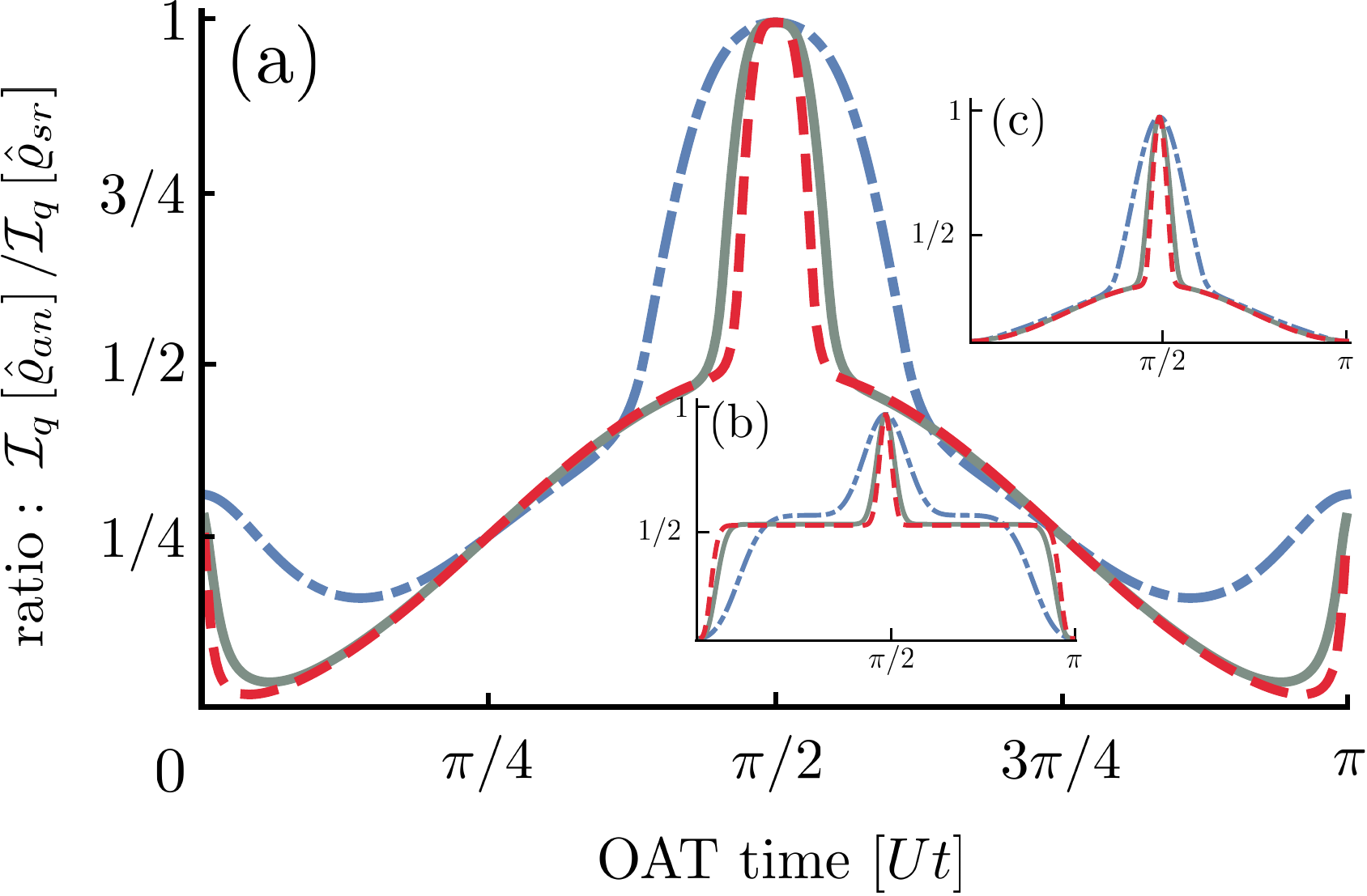}
    \caption{(a): The main figure shows the ratio of the QFI calculated at the antenna and at the source as a function of the OAT time. (b)/(c): The QFI at the source/antenna, both normalized
      to the Heisenberg limit $M^2$. The curves are for $M=10$ (dot-dashed blue), $M=50$ (solid green) and $M=100$ (dashed red).}
    \label{fig.oat}
\end{figure}

This is the central result of our work---a careful design of two-qubit interactions allows a complete transfer of information from the source, passing through a many-body medium, 
to the antenna. Crucially, the 
Heisenberg scaling is fully preserved and can now be accessed by simple measurements on a single receiving qubit. Another important implication of these considerations is that
coupling an $M$-body source to just a single qubit (with no other particles in the medium) would give the same results as in Eqs~\eqref{eq.sn} and~\eqref{eq.heis.2}.

The GHZ state as in Eq.~\eqref{eq.ghz} can be generated by the One-Axis Twisting (OAT) procedure, which takes the source in a product state $\ket{+1}_x^{\otimes M}$ and
acts on it with an Ising-type Hamiltonian as in Eq.~\eqref{eq.ising} with the ATA coupling $U$ between all pairs of source qubits~\cite{pezze2009entanglement}. 
The OAT first squeezes the source and at the optimal time $Ut=\pi/2$ creates
the GHZ state. Thus, the OAT is a good way to generate source states of different entanglement strength \cite{Plodzien2022} (by varying the parameter $Ut$)  and test what fraction of the
information encoded in such states through the transformation~\eqref{eq.rot} reaches the antenna. In Fig.~\ref{fig.oat} (a) we show the ratio of the QFI calculated 
at the antenna to that computed at the source at different instants of the OAT procedure with $M=10$, 50 and 100. Crucially, the figure shows that this ratio is large at most times. Thus. the majority of
the signal reaches the antenna even for moderately entangled source states. In the Appendix we present a plot similar to Fig.~\ref{fig.oat} but obtained using non-zero $\theta$.

\subsection{Optimal measurement}

The QFI is the maximum amount of information that can be extracted from any measurements made on the system. For a given observable $\hat A$ the amount of information that can be extracted
from its values $a_i$ is 
\begin{align}
  \Ic{\hat\varrho_{an}}=\sum_i\frac1{p(a_i|\theta)}\left(\frac{\partial p(a_i|\theta)}{\partial\theta}\right)^2,
\end{align} 
where the probability $p(a_i|\theta)$ is given by the positive-defined measurement operator $\hat\Pi(a_i)$, $p(a_i|\theta)=\tr{\hat\varrho_{an}(\theta;t)\hat\Pi(a_i)}$
with $\sum_i\hat\Pi(a_i)=\hat{\mathds1}$. If the measurement on the receiving qubit is  performed in the $y$-basis, so that
$p(a_{1/2}|\theta)$ are the probabilites of finding the qubit in  $\ket{\pm1}_y^{(an)}$, then using the expression for the density matrix from Eq.~\eqref{eq.dens}
and working around $\theta=0$ we get~\cite{supp}
\begin{align}
  \Ic{\hat\varrho_{an}}=4\modsq{\dot a}
\end{align} 
which equals the QFI from Eqs~\eqref{eq.qfi.gen} and~\eqref{eq.heis.2}. Thus, for both separable and entangled states we have identified the optimal measurement that recovers full information 
about $\theta$ from a single-qubit antenna.

\subsection{Fine-tunning}

Naturally, the scheme presented here requires fine tuning of the interaction parameters. Otherwise the sine and cosine functions will oscillate out-of-phase and degrade the signal. 
Therefore, smaller chains, e.g. where  a single qubit receives the information from an $M$-qubit source in the absence of other qubits forming the medium, 
would be easier to realize. Also, the optimal times need to be correctly targeted. For example, a product of $2\mu$ cosine functions oscillating in phase, as in Eq.~\eqref{eq.qfi.gen}, can be approximated by 
$\cos^{2\mu}(2J_2t)\simeq e^{-2\mu(2J_2t-m\pi)^2}$ with $m\in\mathbb N$.
Thus the signal decreases exponentially as $t$ deviates from the optimal value. 

\subsection{Entanglement-depth certification}

Before we finish, we note the possibility of using this protocol to certify the entanglement depth of the source. Namely, when $\Id{\hat\varrho_{sr}}/M\ge k$, then source has $k$-depth entanglement ~\cite{Hyllus2012,Toth2012}. To experimentally obtain $\Id{\hat\varrho_{sr}}$ directly from the source, 
a set of measurements of collective spin operators $\hat{S}_\alpha$,  and $\hat{S}_\alpha\hat{S}_\beta$,  $\alpha, \beta = x,y,z$, is necessary, which is a non-trivial task from the experimental point of view. 
However, because our protocol allows for a full quantum information transfer from the source to the antenna, 
such an entanglement-depth certification can be done via single qubit quantum state tomography performed on the latter.

\section{Conclusions}

In this work we have shown that it is possible to transfer  information from a many-body source to an antenna almost losslessly in a spin-$1/2$ chain. 
The signal traverses a multi-qubit medium
and the dynamics is generated by an Ising-type Hamiltonian. While for an $M$-body source forming a separable state the information reaching the antenna is slightly reduced, 
it is possible to transfer a complete signal either for $M=1$ or with an $M$-qubit GHZ state. It is the latter result we find most remarkable---simple single-qubit measurements on the antenna, 
which we identify, allow to determine the value of the parameter with Heisenberg-limited precision. 
This protocol also allows the remote certification of an entanglement depth of the source using the QFI  
of a single-qubit antenna~\cite{Hyllus2012, Toth2012}. We believe that the method discussed here could improve the precision of quantum devices and simplify metrological protocols. 

\section*{Acknowledgements}

This work was supported by the National Science Centre, Poland, within the QuantERA II Programme that has received funding from the European Union’s Horizon 2020 
research and innovation programme under Grant Agreement No 101017733, Project No. 2021/03/Y/ST2/00195.
M.P. acknowledges support from: European Research Council AdG NOQIA; MCIN/AEI (PGC2018-0910.13039/501100011033, CEX2019-000910-/10.13039/501100011033, Plan National FIDEUA PID2019-106901GB-I00, Plan National STAMEENA PID2022-139099NB, I00, project funded by MCIN/AEI/10.13039/501100011033 and by the “European Union NextGenerationEU/PRTR" (PRTR-C17.I1), FPI); QUANTERA DYNAMITE PCI2022-132919, QuantERA II Programme co-funded by European Union’s Horizon 2020 program under Grant Agreement No 101017733;
Ministry for Digital Transformation and of Civil Service of the Spanish Government through the QUANTUM ENIA project call - Quantum Spain project, and by the European Union through the Recovery, Transformation and Resilience Plan - NextGenerationEU within the framework of the Digital Spain 2026 Agenda;
Fundació Cellex; Fundació Mir-Puig; Generalitat de Catalunya (European Social Fund FEDER and CERCA program; Funded by the European Union. Views and opinions expressed are however those of the author(s) only and do not necessarily reflect those of the European Union, European Commission, European Climate, Infrastructure and Environment Executive Agency (CINEA), or any other granting authority. Neither the European Union nor any granting authority can be held responsible for them (HORIZON-CL4-2022-QUANTUM-02-SGA PASQuanS2.1, 101113690, EU Horizon 2020 FET-OPEN OPTOlogic, Grant No 899794, QU-ATTO, 101168628), EU Horizon Europe Program (This project has received funding from the European Union’s Horizon Europe research and innovation program under grant agreement No 101080086 NeQSTGrant Agreement 101080086 — NeQST); ICFO Internal “QuantumGaudi” project.

The data that support Fig.~\ref{fig.oat} are openly available~\cite{data}.

\appendix

\section{General expression for the antenna's density matrix}\label{sec.gen}

The general expression for the $N$-qubit density matrix after the phase-imprint is
\begin{align}
  \hat\varrho(\theta)&=\sum_{\vec s,\vec s'=\pm1}\varrho_{\vec s,\vec s'}e^{-i\hat H_{sr}\theta}\ketbra{\vec s}{\vec s'}e^{i\hat H_{sr}\theta}=\nonumber\\
  &=\sum_{\vec s,\vec s'=\pm1}\varrho_{\vec s,\vec s'}(\theta)\ketbra{\vec s}{\vec s'}.
\end{align}
The subsequent time evolution gives
\begin{align}\label{SM.dens.gen}
    \hat\varrho(\theta;t)=\sum_{\vec s,\vec s'=\pm1}\varrho_{\vec s,\vec s'}(\theta)e^{-it\sum_{i>j=1}^NJ_{ij}(s_is_j-s_i's_j')}\ketbra{\vec s}{\vec s'}.
\end{align}
The next step is to trace out all the degrees of freedom apart from that related to the antenna, here labeled with an index $i_{an}$. 
For the diagonal term of the antenna's density matrix all indices are set equal, namely $\vec s=\vec s'$ hence the diagonal does not change, giving $\varrho^{(+1,+1)}_{an}(\theta,t)=p$
and $\varrho^{(-1,-1)}_{an}(\theta,t)=1-p$ and the value of $p$ is given by the initial condition. 

For the off-diagonal term, denoted in the main text by $a$, indices are $\vec s=\vec s'$ for all
qubits apart from the antenna. Since for the antenna $s_{i_{an}}=+1$ and $s'_{i_{an}}=-1$ (or vice-versa for the other off-diagonal term), then the time-dependent exponent becomes
\begin{align}
  e^{-it\sum\limits_{i>j=1}^NJ_{ij}(s_is_j-s_i's_j')}\ \longrightarrow\ e^{-2it\sideset{}{'}\sum\limits_{i=1}^NJ_{i,i_{an}}s_i}.
\end{align}
Only those terms contribute to the sum, where one of the indices points to the antenna. The other terms cancel out (due to the trace). The prime informs that the sum runs through 
all the indices apart from $i_{an}$. Analogical argument applies to the external sum in Eq.~\eqref{SM.dens.gen}, while the matrix element $\varrho_{\vec s,\vec s'}(\theta)$ becomes
$\tilde\varrho_{\vec s,\vec s}(\theta)$, where the tilde denotes that again all the indices are set pairwise equal apart from $s_{i_{an}}$ and $s'_{i_{an}}$. This justifies the expression used in the main text.

\section{Off-diagonal element $a$: specific cases}

We now calculate the off-diagonal element of the antenna's density matrix for a separable and entangled state.

\subsection{Separable state}

First we assume that the full chain initially is in a product of $\ket{+1}_x$ states. By taking the phase transformation to be, for instance, in the form
\begin{align}
  \hat H_{sr}=\frac12\sum_{i_{sr}=1}^M\hat\sigma_y^{(i_{sr})}.
\end{align}
we note that each single qubit of the source undergoes the following transformation
\begin{widetext}
  \begin{align}\label{SM.coeff}
    e^{-\frac i2\theta\hat\sigma_y}\ket{+1}_x&=\left[\hat{\mathds1}\cos\left(\frac\theta2\right)-i\sin\left(\frac\theta2\right)\hat\sigma_y\right]\frac1{\sqrt2}\big({\ket{-1_0}+\ket{1_0}}\big)=\nonumber\\
    &=\frac1{\sqrt2}\ket{-1}_z\left[\cos\left(\frac\theta2\right)+\sin\left(\frac\theta2\right)\right]+\frac1{\sqrt2}\ket{+1}_z\left[\cos\left(\frac\theta2\right)-\sin\left(\frac\theta2\right)\right]=\nonumber\\
    &=\frac1{\sqrt2}\sum_{s=\pm1}\left[\cos\left(\frac\theta2\right)+(-1)^{\frac{s+1}2}\sin\left(\frac\theta2\right)\right]\ket{s}_z.
  \end{align}
\end{widetext}
Hence the complete state after the transformation has the form
\begin{align}
  \ket{\psi(\theta)}=\frac{1}{2^{N/2}}\sum_{\vec s}C(\vec s)\ket{\vec s},
\end{align}
where 
\begin{align}
  C(\vec s)=\prod_{i=1}^Nc_i(\theta)
\end{align}
and $c_i=1$ for non-source qubits, while for the $M$ source qubits, the single qubit coefficient is given by Eq.~\eqref{SM.coeff}. The time evolution imprints the phase as in Eq.~\eqref{SM.dens.gen}.
With this coefficient at hand, we calculate the off-diagonal element of the density matrix of the antenna. First, consider a part of the sum, where the antenna couples to one of the medium qubits. 
The contribution to the matrix element will be
\begin{align}
  \frac12\sum_{s_i=\pm1}e^{-2itJ_{i,i_{an}}}=\cos(2itJ_{i,i_{an}}).
\end{align}
The coupling to the source qubit will take a different form, namely
\begin{align}
  &\frac12\sum_{s_i=\pm1}e^{-2itJ_{i,i_{an}}}\left[\cos\left(\frac\theta2\right)+(-1)^{\frac{s_i+1}2}\sin\left(\frac\theta2\right)\right]^2=\nonumber\\
  &=\cos(2itJ_{i,i_{an}})+i\sin(2itJ_{i,i_{an}})\sin(\theta).
\end{align}
These two results, combined, give the functions $\mathcal F$ and $\mathcal G$ from the main text.

\subsection{GHZ state}

When the source forms the GHZ state and each of its qubits couples to the antenna with the same strength, it is convenient to use the second quantization, giving the source in the form
\begin{align}
  \ket{\psi_{sr}}=\frac1{\sqrt2}\left(\ket{+}_y+i\ket{-}_y\right),
\end{align}
where
\begin{align}
  \ket{\pm}_y=\frac1{\sqrt{2^{M}M!}}\left(\hat a^\dagger\pm i\hat b^\dagger\right)^M\ket0.
\end{align}
are the minimal and maximal eigen-states of the $\hat J_y=1/(2i)(\hat a^\dagger\hat b-\hat a\hat b^\dagger)$, namely
\begin{align}
  \hat J_y\ket{\pm}_y=\pm\frac M2\ket{\pm}_y.
\end{align}
The source state undergoes a phase-imprint through the $\hat J_y$ rotation, and we obtain
\begin{align}
  \ket{\psi_{sr}(\theta)}=e^{-i\theta\hat J_y}\ket{\psi_{sr}}=\frac1{\sqrt2}\left(\ket{+}_y+ie^{-iM\theta}\ket{-}_y\right).
\end{align}
In order to propagate this state with the Ising Hamiltonian, we need to decompose it in the eigen-states of $\hat J_z=\frac12(\hat a^\dagger\hat a-\hat b^\dagger\hat b)$, namely
\begin{align}
  \ket{\psi_{sr}(\theta)}=\sum_{n=0}^MC_n(\theta)\ket{n,M-n}
\end{align}
with
\begin{align}
  C_n(\theta)=\frac{1}{\sqrt2}\sqrt{\frac1{2^M}\binom Mn}i^{M-n}\left(1+i(-1)^{M-n}e^{-iM\theta}\right).
\end{align}
The Hamiltonian consists of two parts: qubit-qubit coupling within the medium and the collective coupling of the source to the medium qubits
\begin{align}\label{eq.ham.bec}
  \hat H=\sum_{i,j}J_{ij}\hat\sigma_z^{(i)}\hat\sigma_z^{(j)}+\sum_{i}J_i\hat\sigma_z^{(i)}\hat J_z.
\end{align}
The system consits of $M$-body source and $N-M$ medium qubits, each in the $\ket{+}_x$ state, hence the composite state evolves with the Hamiltonian from Eq.~\eqref{eq.ham.bec} giving
\begin{align}
  \ket{\psi(\theta,t)}=&\frac1{2^N}\sum_{\vec s}\sum_{n=0}^MC_n(\theta)e^{-it\sum_{ij}J_{ij}s_is_j}\times\nonumber\\
  &\times e^{-it\sum_{i}J_{i}s_i\left(n-\frac M2\right)}\ket{\vec s}\otimes\ket{n,M-n}.
\end{align}
Just as in the previous case, we construct the density matrix and trace out all the degrees of freedom apart from those of the  $k$-th qubit. The coefficient of the diagonal terms
$\ketbra00$ and $\ketbra11$ will, again, be equal to $1/2$, while the coefficient of the off-diagonal part is
\begin{align}
  a=\frac12\left[\cos^M(\varphi_0)+i^M\sin^M(\varphi_0)\sin(M\theta)\right]\prod_{i}\cos(\varphi_i).
\end{align}
as reported in the main text.

\section{Analytical expression for the QFI}

The antenna's density matrix has the form
\begin{align}
  \hat\varrho_{an}(\theta,t)=\left(\begin{array}{cc}
    \frac12 & a \\
    a^* & \frac 12
  \end{array}\right),
\end{align}
This matrix has the eigenvalues and the corresponding eigen-states equal to
\begin{subequations}
  \begin{align}
    &\lambda_+=\frac12+|a|,\ \ \ket{\psi_+}=\frac1{\sqrt2}\left(e^{i\phi}\ket{0}+\ket1\right),\\
    &\lambda_-=\frac12-|a|,\ \ \ket{\psi_-}=\frac1{\sqrt2}\left(-e^{i\phi}\ket{0}+\ket1\right),
  \end{align}
\end{subequations}
where $\phi$ is the phase of $a$. The QFI is given by 
\begin{align}
  F_Q=2\sum_{i,j=\pm}\frac1{\lambda_i+\lambda_j}\modsq{\bra{\psi_i}\partial_\theta\hat\varrho_{an}(\theta,t)\ket{\psi_j}}.
\end{align}
The derivative of $\hat\varrho_{an}$ is
\begin{align}
  \partial_\theta\hat\varrho_{an}(\theta,t)=\left(\begin{array}{cc}
    0 & a' \\
    (a^*)' & 0
  \end{array}\right).
\end{align}
A straightforward calculation gives
\begin{align}\label{eq.full.app}
  F_q=4\frac{c^2}{1-|a|^2}+4s^2,
\end{align}
where
\begin{subequations}
  \begin{align}
    &c=\frac12(a'e^{-i\phi}+(a^*)'e^{i\phi})\\
    &s=\frac1{2i}(a'e^{-i\phi}-(a^*)'e^{i\phi}).
  \end{align}
\end{subequations}

\section{Classical Fisher information}

We now compute the classical Fisher information, taking as the observable the operator $\hat\sigma_y^{(an)}$. The probabilities of finding the antenna in one of the eigen-states
of this operator are
\begin{align}
  p(\pm1|\theta)=\tr{\hat\varrho_{an}(\theta;t)\hat\Pi_\pm}=\frac12\pm\im a,
\end{align}
where
\begin{align}
  \hat\Pi_\pm=\ketbra{\pm1}{\pm1}_y.
\end{align}
The Fisher information is
\begin{align}
  \Ic{\hat\varrho_{an}}&=\frac1{p(+1|\theta)}\left(\frac{\partial p(+1|\theta)}{\partial\theta}\right)^2\nonumber\\
  &+\frac1{p(-1|\theta)}\left(\frac{\partial p(-1|\theta)}{\partial\theta}\right)^2.
\end{align} 
When working around $\theta=0$, we obtain for all cases $\im a=0$, hence
\begin{align}
  \Ic{\hat\varrho_{an}}=4\modsq{\dot a},
\end{align} 
where the derivative is calculated at $\theta=0$. This is the result used in the main text.

\section{Bi-partite density matrix}

The straightforward calculation for the case of a single-qubit source gives the source-antenna reduced density matrix
\begin{align}
  \hat\varrho_{sr;an}(t)=
  \left(\begin{array}{cccc}
    \frac14 & \alpha & \alpha & \frac14\\
    \alpha^* & \frac14 & \beta & \alpha^*\\
    \alpha^* & \beta & \frac14 & \alpha^*\\
    \frac14 & \alpha & \alpha & \frac14
    \end{array}\right),
\end{align}
where
\begin{align}
  \alpha=\frac14\cos^{N-2}(2t)e^{-it},\ \ \ \beta=\frac14\cos^{N-2}(4t).
\end{align}
This matrix is expressed in the following bi-partite basis: $\ket{-1,-1}_z$, $\ket{-1,+1}_z$, $\ket{+1,-1}_z$, $\ket{+1,+1}_z$ of the Hilbert space $\mathcal H_{sr}\otimes\mathcal H_{an}$.
The partial transpose over, say, antenna's degrees of freedom gives
\begin{align}
  \hat\varrho_{sr;an}^{T_{an}}(t)=
  \left(\begin{array}{cccc}
    \frac14 & \alpha^* & \alpha & \frac14\\
    \alpha & \frac14 & \beta & \alpha^*\\
    \alpha^* & \beta & \frac14 & \alpha\\
    \frac14 & \alpha & \alpha^* & \frac14
    \end{array}\right).
\end{align}
Its four eigen-values are
\begin{subequations}
  \begin{align}
    &\lambda_1(t)=\frac18\left(3+4\beta-\sqrt{(1-4\beta)^2+(16\re{\alpha})^2}\right)\\
    &\lambda_2(t)=\frac18\left(3+4\beta+\sqrt{(1-4\beta)^2+(16\re{\alpha})^2}\right)\\
    &\lambda_3(t)=\frac18\left(1-4\beta-\sqrt{(1-4\beta)^2+(16\im{\alpha})^2}\right)\\
    &\lambda_4(t)=\frac18\left(1-4\beta+\sqrt{(1-4\beta)^2+(16\im{\alpha})^2}\right).
  \end{align}
\end{subequations}
Only $\lambda_3(t)$ can be negative, hence the negativity is equal to
\begin{align}
  \mathcal N(t)=\left|\lambda_3(t)\right|.
\end{align}
Since $(16\im{\alpha})^2=16\Id{\hat\varrho_{an}}$, this justifies the expression used in the main text.

\section{Fidelity of quantum Fisher information transfer}

In the following, we show that preserving the quantum Fisher information unavoidably entails a high
 {process fidelity} with an ideal SWAP channel. 

 The conveyor Hamiltonian acts for time \(t_{\!*}\) on the joint space  
 \(\mathcal H_{\mathrm{sr}}\otimes\mathcal  H_{\mathrm{med}}\otimes
 \mathcal H_{\mathrm{an}}\) with unitary
 \(\hat{U}(t_{\!*})\!=\!\exp(-i\hat{H}t_{\!*})\).
Starting from
\(\hat{\varrho}_{\mathrm{sr}}\!\otimes\!\ket{+}_{\mathrm{med}}\!\bra{+}\!
        \otimes\!\ket{+}_{\mathrm{an}}\!\bra{+}\),
the antenna output is
\begin{align}
  \hat \varrho_{\mathrm{an}}^{\;\text{out}}
    = \operatorname{Tr}_{\mathrm{sr},\mathrm{med}}
      \bigl[\hat{U}(t_{\!*})\,(\varrho_{\mathrm{sr}}\otimes\ket{0}\!\bra{0}_{\mathrm{med}}\!\otimes\!
      \ket{0}\!\bra{0}_{\mathrm{an}})\,\hat{U}^{\dagger}(t_{\!*})\bigr].
\end{align}
Packaging the map \(\hat{\varrho}_{\mathrm{sr}}\!\mapsto\!\hat \varrho_{\mathrm{an}}^{\;\text{out}}\) yields the CPTP
 {conveyor channel}
\begin{align}
 \displaystyle
    \mathcal T:\;\hat\varrho_{\mathrm{sr}}\;\longmapsto\;\hat\varrho_{\mathrm{an}}^{\;\text{out}}  .
\end{align}
Perfect behaviour would deposit the sensor state  {unchanged} onto
the antenna qubit:
\begin{align}
 \displaystyle
    \mathcal U_{\text{swap}}:\;
      \hat \varrho_{\mathrm{sr}}\;\longmapsto\;\hat \varrho_{\mathrm{an}}=\hat \varrho_{\mathrm{sr}} .
\end{align}
For any channel \(\Phi\) the corresponding Choi (Jamiołkowski) matrix is
\begin{align}
  J(\Phi)=(\Phi\otimes\mathbb I)
           \bigl[\ket{\Phi^{+}}\!\bra{\Phi^{+}}\bigr],\ \ \ 
  \ket{\Phi^{+}}=\tfrac1{\sqrt d}\sum_{j=1}^{d}\ket{j}_{A}\ket{j}_{B},
\end{align}
with \(d=2^{M}\!:=\!\dim\mathcal H_{\mathrm{sr}}\).

The standard way to compare two channels $\Phi$ and $\Psi$ is to compare
their  {Choi} matrices
$J(\Phi)=(\Phi\!\otimes\!\mathbb I)\ket{\Phi^{+}}\!\bra{\Phi^{+}}$
and compute the Uhlmann fidelity
$F_{\mathrm{state}}(J(\Phi),J(\Psi))$; we call the resulting
\begin{align}
   \mathcal F_{\mathrm p}=F_{\mathrm{state}}
          \bigl(J(\mathcal T),J(\mathcal U_{\mathrm{swap}})\bigr) = \Bigl\|\sqrt{J(\mathcal T)}\sqrt{J(\mathcal U_{\text{swap}})}
        \Bigr\|_{1}^{2}
\end{align}
the  {process fidelity}.  
It lies between 0 and 1 and equals~1 only if
$\mathcal T$ and $\mathcal U_{\mathrm{swap}}$ are
indistinguishable on {all} inputs.
After imprinting a  phase
\(\theta\ll1\) the sensor block is
\begin{align}
    \hat\varrho_{\theta}^{\mathrm{in}}
     = \hat\varrho_{0}
       - i\theta\,[\hat H,\hat\varrho_{0}]
       - \frac{\theta^{2}}{2}\,[\hat H,[\hat H,\hat\varrho_{0}]]
       + O(\theta^{3}).
\end{align}
Because the conveyor channel
\(\mathcal T\) is linear and CPTP map, we get
\begin{align}
  \hat \varrho_{\theta}^{\mathrm{out}}&= \mathcal T(\hat\varrho_{0})- i\theta\,[\mathcal T(\hat H),\mathcal T(\hat \varrho_{0})]\nonumber\\
  &- \frac{\theta^{2}}{2}\,[\mathcal T(\hat H),[\mathcal T(\hat H),\mathcal T(\hat \varrho_{0})]]
  + O(\theta^{3}) .
\end{align}

For an arbitrary density matrix \(\hat\varrho\) and a perturbation
\(\delta\hat\varrho\) the expansion of the fidelity reads
\begin{align}
  F_{\mathrm{state}}(\hat\varrho,\hat\varrho+\delta\hat\varrho)
    = 1 - \tfrac18 \operatorname{Tr}\!\bigl[(\hat{L}_\varrho\,\delta\hat\varrho)^{2}\bigr]
        + O(\delta\hat\varrho^{3}),
\end{align}
where \(\hat L_\varrho\) is the symmetric logarithmic derivative (SLD).
{Applying (S3) with
\(\hat\varrho=\hat\varrho_{\theta}^{\mathrm{in}}\) and
\(\delta\hat\varrho\hat=\varrho_{\theta}^{\mathrm{out}}-\hat\varrho_{\theta}^{\mathrm{in}}\)}
and keeping only the \(\theta^{2}\) terms gives
\begin{align}
  &1-F_{\mathrm{state}}\bigl(\hat\varrho_{\theta}^{\mathrm{in}},\hat\varrho_{\theta}^{\mathrm{out}}\bigr)\\
  &= \frac{\theta^{2}}{4}\,
      \Bigl|
    \operatorname{Tr}\!\bigl(\hat\varrho_{0}\hat L_{\mathrm{in}}^{2}\bigr)
        -\operatorname{Tr}\!\bigl(\mathcal T(\hat\varrho_{0})\hat L_{\mathrm{out}}^{2}\bigr)
      \Bigr|
      +O(\theta^{3}).\nonumber
\end{align}

But
\(\operatorname{Tr}(\hat\varrho_{0}\hat L_{\mathrm{in}}^{2})=F_{Q}^{\mathrm{in}}\) and
\(\operatorname{Tr}(\mathcal T(\hat\varrho_{0})\hat L_{\mathrm{out}}^{2})=F_{Q}^{\mathrm{out}}\) are the Quantum Fisher Information, thus we obtain
\[
    1-F_{\mathrm{state}}
     \bigl(\hat\varrho_{\theta}^{\mathrm{in}},\hat\varrho_{\theta}^{\mathrm{out}}\bigr)
      = \frac14\,
        \bigl|F_{Q}^{\mathrm{in}}-F_{Q}^{\mathrm{out}}\bigr|\,
        \theta^{2}
        + O(\theta^{3}).
\tag{S4}
\]

\noindent
If the conveyor preserves QFI exactly
\(\bigl(F_{Q}^{\mathrm{in}}=F_{Q}^{\mathrm{out}}\bigr)\),
the first non-vanishing term is cubic in \(\theta\); hence
input and output states remain practically indistinguishable throughout, and process fideity is ${\cal F}= 1$.

As such, preserving the quantum Fisher information {automatically} guarantees a high
process fidelity to the ideal SWAP channel.

\section{Experimental overhead for extracting Quantum Fisher Information from $M$-qubit GHZ state}

To demonstrate that any protocol truly reaches the quantum‐Fisher‐information
(QFI) predicted in theory, one must characterise the output state
and the measurement.  In practice this means performing (at least
partial) quantum state tomography or, equivalently, estimating enough
expectation values to compute a tight lower bound on QFI. To see why measurement on single qubit antenna is beneficial comparing to M-qubit GHZ measurement, we discuss three distinct aspects:

\medskip
\textbf{Tomographic overhead.}
A general $M$–qubit state is specified by an exponentially growing number of parameters; even sparsity‐aware techniques
(compressed sensing, permutational symmetries) still require a number of
measurement settings that grows \(\gtrsim \mathrm{poly}(M)\).
By contrast, a  {single} qubit is fully specified by just three real
numbers (\(\langle\sigma_x\rangle,\langle\sigma_y\rangle,\langle\sigma_z\rangle\)).
The experimental burden therefore scales as

\[
  \#\text{settings}\;=\;
  \begin{cases}
    \mathcal O(M^2) & \text{\footnotesize(GHZ tomography },\\[4pt]
    3 & \text{\footnotesize(antenna qubit)}.
  \end{cases}
\]

\textbf{Sample complexity.} For a fixed target precision \(\delta\!\varrho\), the required number of experimental shots obeys \(N_{\rm shots}\propto (\#\text{settings})/\delta^2\).
Rplacing  \(M\!>\!10\) qubits with a one-qubit antenna reduces both the measurement settings and the total sample load by at least an order of magnitude, sometimes two.

\textbf{Calibration of optimal observables.} In classical estimation theory  the ``score'' is the derivative of the log-likelihood
$s(x|\theta)=\partial_\theta\ln p(x|\theta)$, and its variance is the classical Fisher information. 
In quantum theory, the probability distribution $p(x|\theta) = \operatorname{Tr}[\hat \varrho_\theta \hat \Pi_x]$ depends on the density matrix $\hat \varrho_\theta$ and on the POVM $\{\hat  \Pi_x\}$. 
For a quantum probe $\hat \varrho_\theta$, the score is defined via the  symmetric logarithmic derivative (SLD), 
through the  \textit{score operator} $\hat L_\theta$, which defines score for quantum system, i.e. 
$\partial_\theta\hat \varrho_\theta=\tfrac12\bigl(\hat \varrho_\theta \hat L_\theta+\hat L_\theta\hat \varrho_\theta\bigr)$,
where \(\hat L_\theta\) is Hermitian operator. 

This is the quantum analogue of the derivative of the log-likelihoood; when the measurement is prepared in the eigenbasis of $L_\theta$ its eigenvalues play the role of the classical score. 
The QFI is defined as the variance of the SLD, i.e. $    F_Q(\theta)=\operatorname{Tr}\!\bigl(\hat \varrho_\theta \hat L_\theta^{2}\bigr)$, 
and upper-bounds the classical Fisher information
$F_C\le F_Q$ for {any} POVM.
Projecting onto the eigenbasis of $\hat L_\theta$ {saturates} this bound,
achieving the quantum Cramér–Rao limit
$\operatorname{Var}(\hat\theta)\ge 1/(mF_Q)$. To construct the SLD one must diagonalise the density matrix $\hat \varrho_\theta=\sum_j\lambda_j\ket{j}\!\bra{j}$, and then
\begin{equation}
    \bra{j }\hat L_\theta\ket{k}=
      \begin{cases}
        \dfrac{2\,\bra{j}\partial_\theta\hat \varrho_\theta\ket{k}}
                   {\lambda_j+\lambda_k}, & \lambda_j+\lambda_k\neq0,\\[8pt]
        0, & \lambda_j+\lambda_k=0.
      \end{cases}
  \label{eq:SLD-mixed}
\end{equation}
When considering pure state  $\hat \varrho_\theta=\ket{\psi_\theta}\!\bra{\psi_\theta}$ with
$\langle\psi_\theta|\partial_\theta\psi_\theta\rangle=0$ one gets
\begin{subequations}\label{eq:SLD-pure}
  \begin{align}
    \hat   L_\theta = 2\Bigl(\ket{\partial_\theta\psi_\theta}\!\bra{\psi_\theta} + \ket{\psi_\theta}\!\bra{\partial_\theta\psi_\theta}\Bigr),\\
    F_Q = 4\Bigl(\langle\partial_\theta\psi_\theta|\partial_\theta\psi_\theta\rangle -
    |\langle\psi_\theta|\partial_\theta\psi_\theta\rangle|^{2}
  \Bigr).
  \end{align}
\end{subequations}

The SLD operator, and corresponding QFI for single antenna qubit and M-qubit GHZ state are summarized in the following table:
\begin{widetext}
  \begin{center}\small
    \begin{tabular}{lccc}
      \hline\hline
      Probe state & $\partial_\theta\varrho_\theta$ & $L_\theta$ & $F_Q$ \\
      \hline
      Single qubit, $z$–rotation &
      $\tfrac12(-\sin\theta\,\sigma_x+\cos\theta\,\sigma_y)$ &
      $\sigma_y$ & $1$ \\[4pt]
      $M$–qubit GHZ
      $\displaystyle\frac{\ket{0}^{\otimes M}+e^{iM\theta}\ket{1}^{\otimes M}}{\sqrt2}$ &
      $\dfrac{iM}{2}
      (\ket{0}^{\otimes M}\!\bra{1}^{\otimes M}-\text{h.c.})$ &
      $M\sigma_y^{\otimes M}$ & $M^{2}$\\
      \hline\hline
    \end{tabular}
  \end{center}
\end{widetext}
The antenna-qubit protocol needs only a
{single} Pauli rotation (\(\hat \sigma_y\)),
whereas the direct GHZ read-out requires the $M$–body parity
$\hat \sigma_y^{\otimes M}$.

Now, let us focus on the sensitivity of the protocol to the calibration errors for the single-qubit antenna with and $M$-qubit GHZ state. Let every qubit be measured along
$\hat \sigma_{y,\varepsilon}=
  \hat \sigma_y\cos\varepsilon+\hat \sigma_x\sin\varepsilon$
due to a phase slip $\varepsilon$ in the intended $\pi/2$ pulse. For a single antennna qubit, having the outcome probabilities
$p_\pm=\tfrac12[1\pm\cos\varepsilon\,\theta]$
the classical Fisher information is
\(
  F_C^{(1)}=\cos^{2}\!\varepsilon\,F_Q^{(1)}.
\) On the other hand for an $M$–qubit GHZ (identical slip on all qubits), the measured observable is
$O=(\hat \sigma_{y,\varepsilon})^{\otimes M}$.
For $\theta\!\ll\!1$:
\[
  \langle\hat  O\rangle\simeq
  M\theta\,\cos^{M}\!\varepsilon,
  \qquad
  \operatorname{Var}(\hat O)=1-\langle \hat O\rangle^{2}\simeq1,
\]
hence
\[
  F_C^{(\mathrm{GHZ})}
     =\bigl[\partial_\theta\langle \hat O\rangle\bigr]^{2}
     =M^{2}\cos^{2M}\!\varepsilon
     =F_Q^{(\mathrm{GHZ})}\,e^{-M\varepsilon^{2}}.
\]

The phase mismatch at the level of \(\varepsilon=1^{\circ}\) only slightly degrades
$F_C/F_Q$, however when considering M-qubit GHZ state, then ratio $F_C/F_Q$ vanishes exponentially with  $M$. As such, it is beneficial to prepare single qubit measurement, comparing to M-qubit GHZ measurement.

\section{The QFI for $\theta\neq0$}

\begin{figure}[t!]
    \centering
    \includegraphics[width=1.0\linewidth]{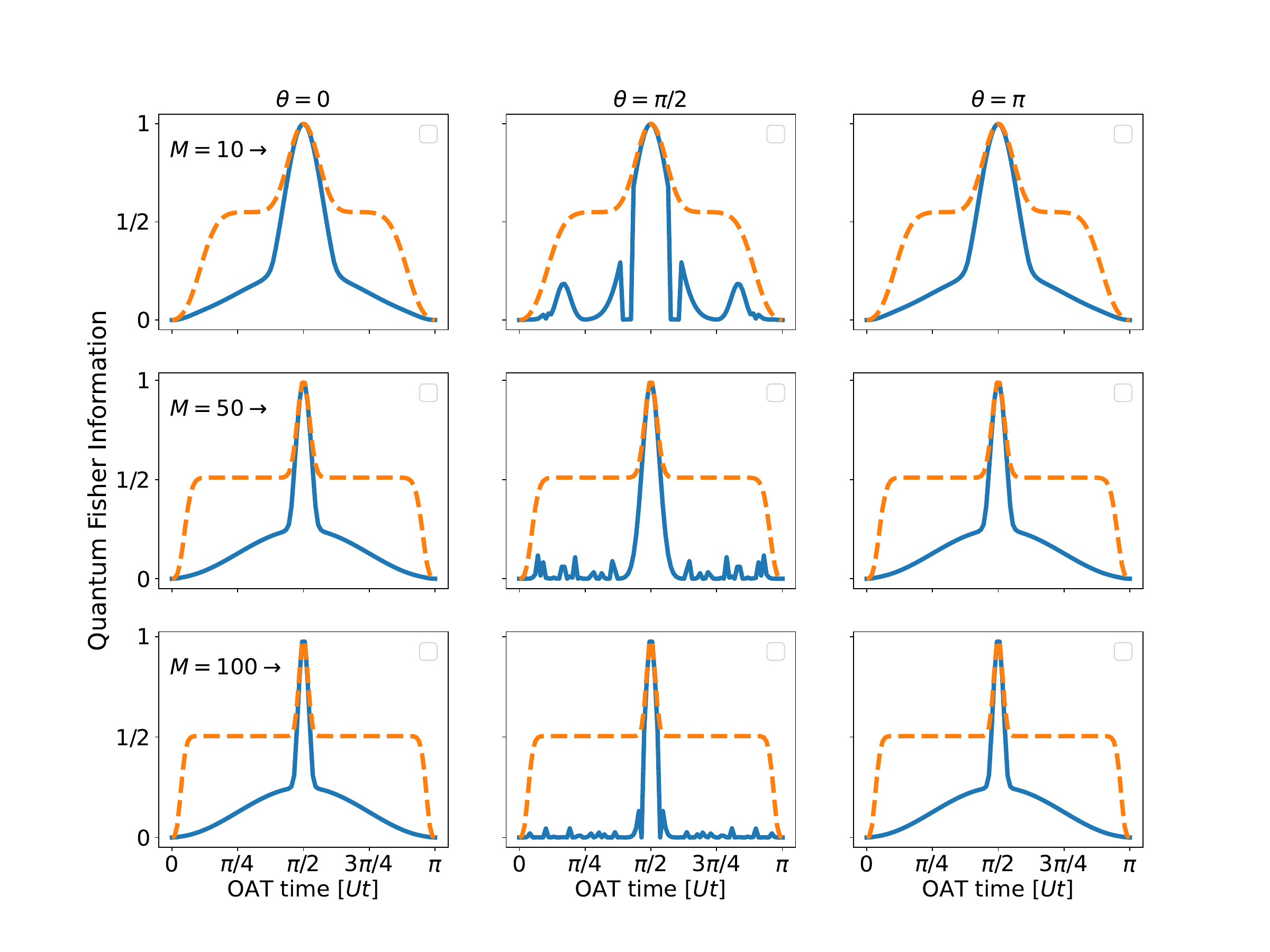}
    \caption{The QFI calculated at the antenna and at the source as a function of the OAT time for $M=10$, $M=50$ and $M=100$ (rows) and with $\theta=0$, $\pi/2$ and $\pi$ (columns).}
    \label{fig.oat.supp}
\end{figure}

In this section, we compare the QFI at the antenna to the readout at the $N$-qubit OAT state for values of the parameter of $0,\ \pi/2,$ and $\pi$ with $M = 10, 50,$ and 100, see Fig.~\ref{fig.oat.supp}. 
Although the transfer can be sub-optimal for the non-zero values of the parameter, keep in mind that, in many experiments, $\theta$ is small and roughly known. 
This allows the experimentalists to add an external field that effectively shifts the value of $\theta$ close to zero.

\end{document}